# Local strain inhomogeneities during the electrical triggering of a metal-insulator transition revealed by the x-ray microscopy

*Pavel Salev[1], Elliot Kisiel[2,3], Dayne Sasaki[4], Brandon Gunn[2], Wei He[2], Mingzhen Feng[4], Junjie Li[2], Nobumichi Tamura[5], Ishwor Poudyal[3], Zahir Islam[3], Yayoi Takamura[4], Alex Frano[2], Ivan K. Schuller[2]*

[1]Department of Physics and Astronomy, University of Denver, Denver CO
[2]Department of Physics, University of California San Diego, La Jolla CA
[3]X-ray Science Division, Argonne National Laboratory, Argonne IL
[4]Department of Materials Science and Engineering, University of California Davis, Davis CA
[5]Advanced Light Source, Lawrence Berkeley National Laboratory, Berkeley CA

**Electrical triggering of a metal-insulator transition (MIT) often results in the formation of characteristic spatial patterns such as a metallic filament percolating through an insulating matrix or an insulating barrier splitting a conducting matrix. When the MIT triggering is driven by electrothermal effects, the temperature of the filament or barrier can be substantially higher than the rest of material. Using x-ray microdiffraction and dark-field x-ray microscopy, we show that electrothermal MIT triggering leads to the development of an inhomogeneous strain profile across the switching device, even when the material does not undergo a 1st order structural phase transition coinciding with the MIT. Diffraction measurements further reveal evidence of lattice distortions and twinning occurring within the MIT switching device, highlighting a qualitative distinction between the electrothermal process and equilibrium thermal lattice expansion in nonlinear electrical systems. Electrically induced strain development, lattice distortions, and twinning could have important contributions in the MIT triggering process and could drive the material into non-equilibrium states, providing an unconventional pathway to explore the phase space of strongly correlated electronic systems.**

## I. Introduction

Applying voltage to metal-insulator transition (MIT) materials can trigger a large resistance change producing volatile resistive switching [1,2]. MIT switches are actively pursued for practical applications, such as selectors in crossbar memory arrays [3,4], rf and optical switches [5–7], and spiking devices for neuromorphic and stochastic computing [8–13]. The physical mechanism that drives the electrical MIT triggering remains a subject of debates. Under an electric stimulus, both Joule heating and electrostatic charge doping can contribute to the MIT triggering [14,15]. Because of the intricate coupling between charge, spin, lattice, and orbital degrees of freedom in MIT materials [16–21], experimental studies providing diverse information about the electrically induced changes in the electronic, magnetic and structural order are necessary to establish the basic understanding of the MIT triggering process.

Some MIT materials, such as $VO_2$ and $V_2O_3$, have a 1st order structural transition coinciding with the electronic transition [22]. Transmission electron microscopy, micro- and nanodiffraction measurements in $VO_2$ switching devices have shown that, when an electric stimulus triggers the MIT, the corresponding structural transition is also induced, which results in the formation of a rutile metal filament inside a monoclinic insulator matrix [23–25]. Other materials, for example, $V_3O_5$, $NbO_2$, $SmNiO_3$, and $(La,Sr)MnO_3$ (LSMO), do not undergo a 1st order structural transition coinciding with the MIT under equilibrium conditions. The crystal structure evolution under strong electric stimuli in such materials remains unknown. Because the MIT can be triggered at temperatures hundreds of Kelvins below the transition temperature ($T_c$) [26–28], it can be expected that thermal gradients associated with the electrically induced local phase transition can induce a substantial thermal expansion of the crystal lattice, altering the local strain. It is well established that MIT materials are highly susceptible to crystal lattice deformations [29–32]; therefore, the development of a local strain during the electrical switching may play an important role in the MIT triggering process.

In this work, we explored the electrically driven crystal structure evolution in LSMO switching devices. LSMO is a particularly interesting system because it exhibits a metal-to-insulator switching [33–35], in contrast to the more common insulator-to-metal switching observed in materials such as $VO_2$, $V_2O_3$, $V_3O_5$, $NbO_2$, $SmNiO_3$, *etc*. The MIT triggering in LSMO also drives unusual magnetic phenomena, including the development of uniaxial magnetic anisotropy [34] and anomalous magnetotransport properties [35], whose physical origins are not yet fully understood. By performing *in-operando* x-ray microdiffraction and dark-field x-ray microscopy (DFXM) studies, we observed that electrically driving LSMO across the MIT results in strain development, even though LSMO does not have a 1st order structural transition coinciding with the MIT. The measured strain displayed a spatial profile consistent with the locally induced MIT switching inside the device. We further found that the lattice expansion within the hot regions that underwent the electrically induced MIT switching partially propagates through the entire device, which produces highly inhomogeneous strain and substantial lattice distortions. This work shows that the lattice evolution in the switching devices under applied voltage qualitatively differs from equilibrium thermal behavior, indicating that electrically induced structural changes can be significant even in MIT materials that do not undergo a coincident 1st order structural transition.

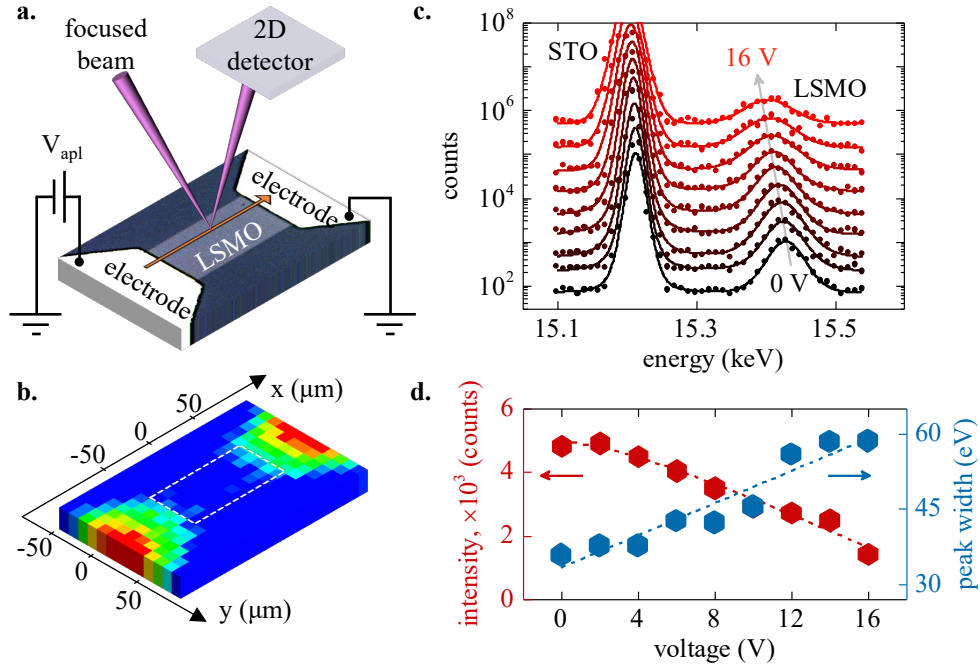

**Fig. 1. a,** Schematic of the x-ray microdiffraction measurement setup. A focused x-ray beam acquires local diffraction patterns at several locations across the LSMO device length. The device is held under a constant applied voltage. The device picture is an optical microscopy image. **b,** X-ray fluorescence map of the LSMO device. The bright regions correspond to the Pd/Au electrodes. The dashed white rectangle indicates the LSMO device location. **c,** Local diffraction patterns acquired at the center of LSMO device under applied voltages in the 0-16 V range at 2 V step. The diffraction patterns were recorded in the vicinity of the (007) STO Bragg peak. The shift of the LSMO peak toward lower energies with increasing voltage indicates the lattice expansion The curves are vertically offset for clarity. **d,** Voltage dependence of the integrated intensity (red symbols) and full width at half maximum (blue symbols) of the LSMO $(007)_{pc}$ Bragg peak. The lines are guides to the eye. All measurements were performed at 180 K.

## II. X-ray microdiffraction

We explored the strain development during electrical MIT triggering in two-terminal $La_{0.7}Sr_{0.3}MnO_3$ devices (Fig. 1a). The devices were patterned in a 50-nm-thick film epitaxially grown on a $SrTiO_3$ (STO) substrate [36,37]. LSMO is a ferromagnetic metal at low temperatures and it transitions into a paramagnetic insulator above $T_c \sim$ 340 K [38]. Applying voltage to LSMO can trigger the MIT, even at the temperatures that are hundreds of Kelvin below $T_c$ [33]. This electrical MIT triggering is mediated by Joule heating, and it produces volatile resistive switching, i.e., a switching that persists while the voltage is applied but automatically resets when the voltage is turned off. The switching occurs in a characteristic spatial pattern: the formation of an insulating barrier in a direction perpendicular to the current flow [33,34], in contrast to the more commonly observed filamentary percolation [39–41]. Even though the MIT in LSMO does not involve a 1st order structural transition, thermal gradients associated with the Joule-heating-driven barrier formation can be expected to generate a substantial strain within the device, which we investigate in this work.

We employed x-ray microdiffraction to directly probe the local lattice expansion in the LSMO devices during electrical MIT triggering (Fig. 1a). These measurements were performed at the beamline 12.3.2 of the Advanced Light Source at Lawrence Berkeley National Laboratory. The x-ray beam was focused to a ~5 μm spot using Kirkpatrick-Baez mirrors. Diffraction patterns were acquired at multiple locations along the device length while holding the device under constant voltages in the 0-16 V range. The focused beam was kept fixed at a 45º angle relative to the sample surface and the diffraction data were acquired by scanning the beam energy through a 15.3-16 keV range, which corresponded to the vicinity of the (007) Bragg peak of the STO substrate. The measurements near this higher order Bragg peak helped minimizing the overlap between the STO and LSMO Bragg peaks. The diffracted beam was collected by a 2D detector. The device size was $50\times100$ μm$^2$. Individual switching devices were located using x-ray fluorescence scans by observing the fluorescence from the Pd/Au electrodes within a 9-10 keV range (Fig. 1b). All x-ray microdiffraction measurements were performed at a base temperature of 180 K.

We observed that an applied voltage induces a significant out-of-plane lattice expansion in the LSMO device. Fig. 1c shows the diffraction patterns acquired locally at approximately the center of the device while applying voltages in the 0-16 V range in 2 V steps. As the voltage increases, the LSMO Bragg peak progressively shifts toward lower energies, indicating an increase of the lattice $d$-spacing. Because the LSMO device experiences substantial Joule heating (the maximum dissipated power reached ~300 mW in our experiments), the Bragg peak shift under an applied voltage can be attributed to thermal lattice expansion. We found, however, that there are qualitative differences between the voltage / Joule heating induced lattice expansion seen here as compared to equilibrium thermal properties. Under an applied voltage, the integrated Bragg peak intensity decreases rapidly, by a factor of ~3.5 at 16 V (Fig. 1d, red line). Simultaneously, the Bragg peak width increases by a factor of ~2 at 16 V (Fig. 1d, blue line). Because Joule heating is expected to substantially increase the device temperature, one might expect that the Debye-Waller factor may account for the peak intensity reduction. However, LSMO Bragg peak intensity and width have a weak temperature dependence in equilibrium (Suppl. Fig. S1). The peak intensity decreases by ~6% and the peak width increases by only 0.4% between 180 K and 400 K at zero voltage. We conclude therefore, that the Debye-Waller factor does not have a significant contribution in our experiments. Under the applied voltage, the observed ~2× increase of the peak width may indicate a spatially inhomogeneous lattice expansion (lateral, i.e., across the device area, and/or vertical, i.e., across the film thickness) at the length-scales below the experimental resolution (~5 μm). As the peak width increase was also accompanied by the ~3.5× decrease of the integrated peak intensity, it is unlikely that such an inhomogeneous lattice expansion is due only to the formation of local regions with different strain. The decrease of the integrated intensity may be due to a change of the lattice form factor, suggesting that the voltage-induced heating may lead to asymmetric in-plane vs. out-of-plane lattice expansion and/or to lattice distortions, thereby reducing the x-ray scattering efficiency.

After observing voltage-induced lattice expansion in the LSMO device center, we next investigate the expansion in different parts of the device. Fig. 2a shows the out-of-plane strain measured at multiple locations across the device length in 10 μm steps while applying constant voltages in the 0-16 V range in 2 V steps. The strain is defined with respect to the out-of-plane $d$-spacing measured at 0 V. Below 8 V, the lattice expansion is the same throughout the device, which is likely due to the uniform Joule heating that precedes the electrothermal MIT triggering [33]. Above 8 V, however, the induced strain is larger in the center region (0.14% at 16 V) than at the device edges (0.09% at 16 V), resulting in bell-shaped strain profiles. This inhomogeneous lattice expansion at higher voltages is consistent with the previous magneto-optical and electrical observations of the insulating barrier formation inside the metallic matrix in LSMO devices during the MIT triggering [33–35]. Because the barrier's resistance is much higher than the metallic matrix resistance, the barrier focuses the dissipated power and experiences strong Joule heating. Larger local Joule heating leads to larger local lattice thermal expansion, as we have observed in the x-ray microdiffraction experiments.

Analysis of the LSMO Bragg peak width and intensity profiles across the device length further shows that the voltage-driven lattice evolution differs in the device center compared to the device edges, supporting the picture that Joule heating is focused inside the switched insulating barrier. At zero voltage, both the peak intensity and width show only small variations with no discernible trend across the device length (dark gray points in Fig. 2 b and c). At 16 V, the intensity and width profiles form bell-shaped curves (red points in Fig. 2 b and c) with the maximum near the device center, identical to the strain maximum location in the lattice expansion profile plot (Fig. 2a). The lower peak intensity and broader peak width at the device center at 16 V suggest that the larger local lattice expansion is accompanied by lattice symmetry distortions, likely due to the strain accommodation, as discussed later in the paper.

The development of strain inhomogeneities across the device length correlates well with the $I$-$V$ characteristics of the LSMO device. The $I$-$V$ curve (Fig. 2d) has two distinct features. The first feature is the onset of negative differential resistance (NDR) at ~8 V, i.e., a part of the curve where the $dV/dI$ slope is negative. Previously, it was found that the NDR onset precedes the MIT triggering and corresponds to the development of a moderate voltage distribution inhomogeneity across the device length [33]. The second important feature in the $I$-$V$ curve is the kink at ~10 V. At this kink, the current decreases quickly with increasing voltage, indicating a rapid increase of the device resistance. In the previous magneto-optical measurements, this kink corresponded to the local phase transition triggering in LSMO and the formation of an insulating barrier inside the device [33]. In the present microdiffraction experiments, we observed that the lattice expansion is the same across the device length (Fig. 2e) up to the NDR onset in the *I-V* curve (~8 V, Fig. 2d). In the NDR region, the voltage dependence of the strain in the device center (red symbols in Fig. 2e) continues to follow approximately the same trend as before the NDR onset. The lattice expansion at the edges, however, slows down (blue symbols in Fig. 2e). The initial departure between the center and edge strain-voltage curves at the NDR onset is consistent with the dissipated power being focused in the device center because of the resistance increase due to the approaching MIT. By considering the strain inhomogeneity defined as $(\varepsilon_{center} - \varepsilon_{edge})/\varepsilon_{edge}$ (Fig. 2f), we observe that this inhomogeneity, emerging at the NDR onset, fully develops, reaching ~40%, at the voltage corresponding to the kink in the *I-V* curve where the spatially uniform metallic state is expected to switch into a phase separated state (insulating barrier splitting the metallic matrix) [33]. Overall, the good correspondence between the strain development and *I-V* curve features shows that electrically induced lattice expansion can be attributed to the applied voltage and dissipated power being focused within the insulating phase barrier when the device undergoes electrothermal MIT switching.

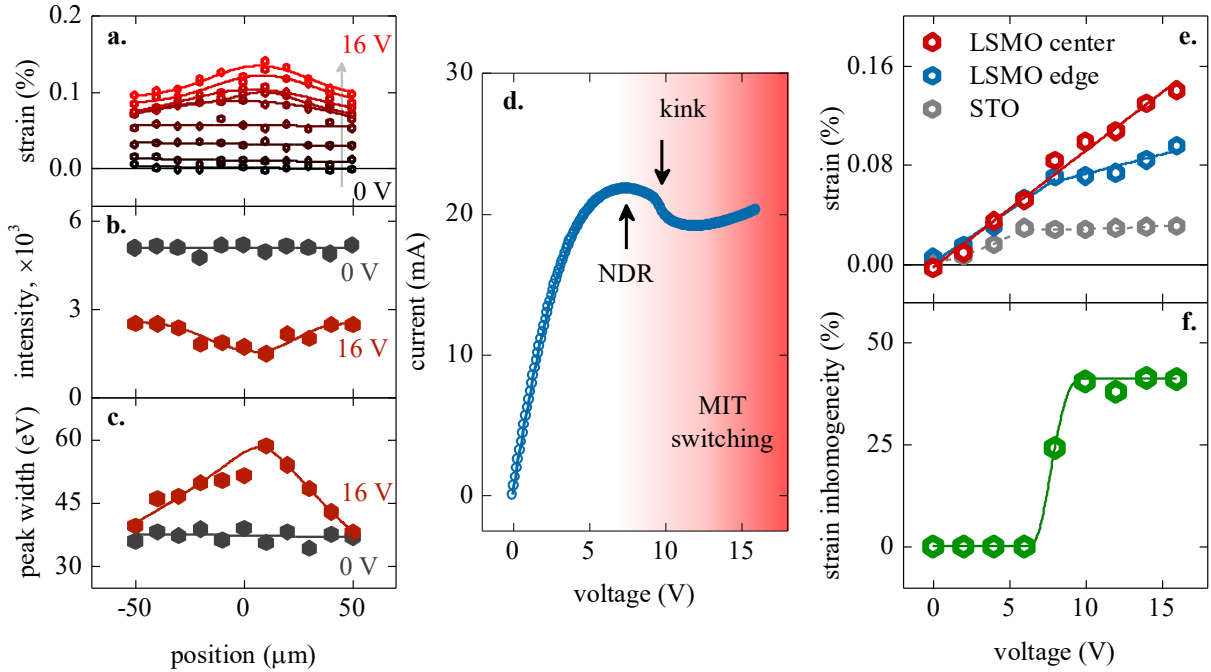

**Fig. 2. a,** Strain profiles across the LSMO device length measured under an applied voltage in the 0-16 V range in 2 V steps. The development of bell-shaped profiles above 8 V, indicating that larger lattice expansion at the device center, can be observed. **b, c,** Integrated intensity (**b**) and width (**c**) of the LSMO $(007)_{pc}$ Bragg peak measured at several locations across the device length while in equilibrium (0 V, dark gray symbols) and above the MIT triggering threshold (16 V, red symbols). The lines are guides to the eye. **d,** I-V curve of the LSMO device exhibiting negative differential resistance (NDR) and a kink at the MIT triggering threshold voltage. **e,** Strain in LSMO at the center (red symbols) and edge (blue symbols) of the device as a function of applied voltage. Gray symbols show strain in the STO substrate averaged over the device length. The lines are guides to the eye. **f,** Strain inhomogeneity that develops in LSMO as a function of applied voltage. The strain inhomogeneity rapidly increases at the NDR onset voltage and reaches a stable value at the MIT triggering threshold. All measurements were performed at 180 K.

While the local temperature increase caused by Joule heating focusing can account for the general trend in the voltage-induced LSMO lattice expansion, several experimental observations highlight the important differences between the electrothermal effects and the equilibrium temperature evolution of the lattice. Electrothermal modeling predicts that the local LSMO temperature reaches a maximum at the barrier formation voltage threshold [33]. Above this threshold, the barrier size grows as the applied voltage increases, but the barrier temperature stays constant. This limiting of the barrier temperature is due to the weak resistance-temperature dependence of LSMO in the insulating phase, which imposes a limit on the dissipated power focusing. The microdiffraction measurements, however, revealed that the lattice at the device center (i.e., in the barrier region) keeps expanding as the applied voltage increases (Fig. 2e). This observed persistent strain increase suggests that local temperature is not the only parameter affecting the lattice constant. It is possible that when the local temperature increases, the associated local expansion can be partially hindered by the adjacent colder regions in the LSMO film and STO substrate enacting mechanical clamping. As the film remains continuous (no visible cracks or mechanical degradation appear in the devices even after millions of switching cycles [33]), the clamping can lead to strain accommodation, which often occurs by spatial modulations in the lattice parameter and/or twinning [42–44].

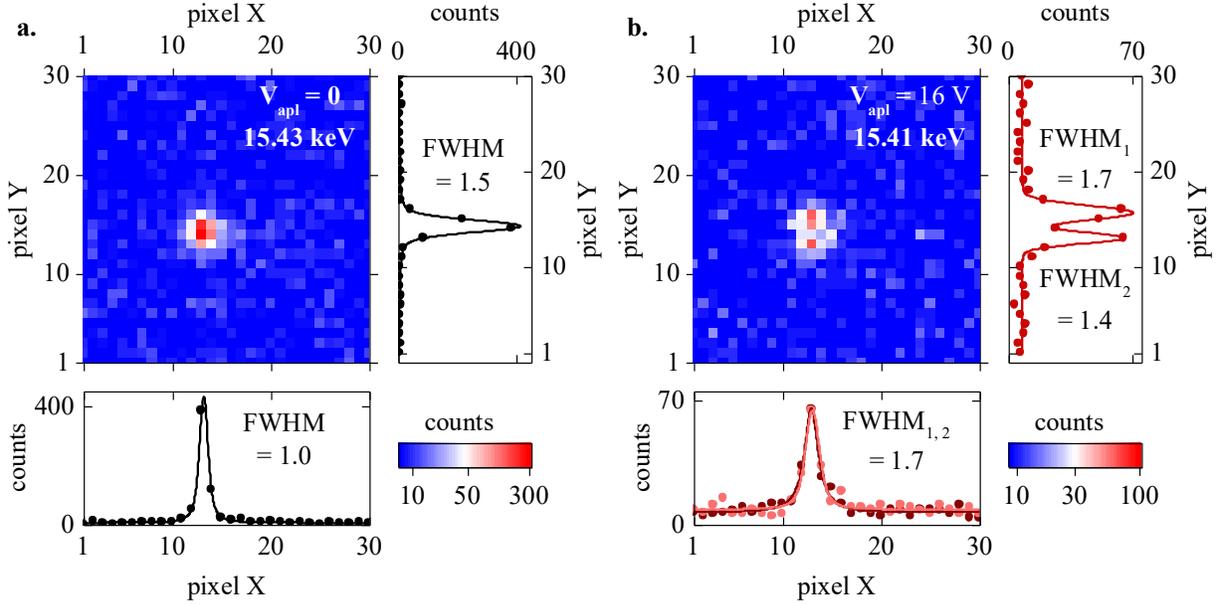

**Fig. 3.** 2D detector images of the LSMO (007)$_{pc}$ Bragg peak acquired in equilibrium (0 V, panel **a**) and above the MIT triggering threshold (16 V, panel **b**). In equilibrium, there is a single diffraction peak indicating structural homogeneity of the LSMO thin film device. Above the MIT triggering threshold, the Bragg peak splits into two well-defined peaks, suggesting that the film breaks into structural twins with different orientations.

Closer examination of the diffraction peaks provides evidence of twin formation in the LSMO devices when the applied voltage is above the MIT triggering threshold. Fig. 3 shows 2D detector snapshots at the x-ray energies corresponding to the LSMO (007)$_{pc}$ Bragg peak. At zero voltage (Fig. 3a), we observed a single diffraction peak, indicating structural homogeneity of the film in equilibrium. At 16 V (Fig. 3a), the peak splits into two well-defined peaks, which suggests that the film breaks into multiple structural domains of slightly different inclinations. The almost even splitting of the diffracted intensity between the two peaks implies that the twin domain size is much smaller compared to the ~5 μm spatial resolution in our microdiffraction experiments. At the twin domain boundaries, the lattice can be highly distorted, which may explain the observed Bragg peak broadening (Fig. 2c) and decreased diffraction intensity (Fig. 2b) when the applied voltage drives LSMO across the phase transition.

### III. Dark-field x-ray microscopy

To further investigate lattice twinning in the switching devices, we employed DFXM, a technique that provides high-resolution, real-space imaging of the diffraction patterns originating from the different crystalline regions in a material [45,46]. The measurements were performed at the beamline 6-ID-C of the Advanced Photon Source at Argonne National Laboratory. Fig. 4a shows the schematics of the DFXM setup. The device area was illuminated by a parallel x-ray beam with an energy of 20 keV and at an incident angle of ~9.2°, which corresponds to the (002)$_{pc}$ LSMO Bragg peak. Imaging was achieved by passing the diffracted beam through a high-efficiency polymeric compound refractive lenses (pCRL) [45,47]. Similar to the lens in an optical microscope, the pCRL refracts the x-ray beam producing a magnified full-field view of the device on the scintillating x-ray detector. The pCRL had a focal length of 131 mm and was placed 140 mm from the sample. The detector was at a distance of 2.3 m from the pCRL. This geometry yielded an x-ray magnification factor of 16×. A 50 μm pinhole and a set of slits in the back focal plane were used to improve the *Q*-space selectivity. An additional optical 5× magnification between the x-ray scintillating detector and optical camera (not shown in Fig. 4a) resulted in an overall 80× magnification. The imaging of the MIT switching in the LSMO device was performed at 115 K to avoid the presence of additional structural features due to the structural phase transition in the STO substrate at 105 K [48]. The LSMO device size was 20×40 μm$^2$, which allowed the entire device area to be illuminated by the x-ray beam without the need to raster. Therefore, our DFXM setup was able to produce a full-field device image in a single exposure, unlike the microdiffraction approach in which the device area was scanned using the focused x-ray beam.

DFXM enabled imaging of small changes in the diffraction conditions when the applied voltage triggered the MIT in LSMO. The *I-V* curve acquired during the x-ray measurements shows a clear discontinuity at 6.5 V (Fig. 4b), which can be attributed to the insulating phase barrier formation [33]. The slightly different threshold voltage compared to the device studied in the microdiffraction experiments is due to the lower base temperature and different device dimensions in the DFXM measurements. The general electrical switching behavior in the two experiments, however, was identical: both I-V curves in Fig. 2d and Fig. 4b display the onset of NDR and a discontinuous kink at the threshold voltage. Fig. 4c shows the measured real-space map of the lattice orientation distribution in the LSMO device at 0 V, i.e., in equilibrium. This map is based on recording a rocking curve around the $(002)_{pc}$ Bragg peak of LSMO and plotting the center of mass (COM) angle for each pixel. It can be observed that there are lattice orientation deviations, approximately ±0.5 mdeg, throughout the device, which could be due to defects formed during the film growth and device fabrication or introduced by the repeated voltage cycling. To highlight the lattice orientation changes induced by the MIT switching, we subtracted this equilibrium orientation distribution map from the maps measured under applied voltages. Thus, Fig. 4 d-f presents the differential maps of the lattice reorientation with respect to the initial state at 9 V, 12 V, and 15 V, i.e., voltages above the switching threshold.

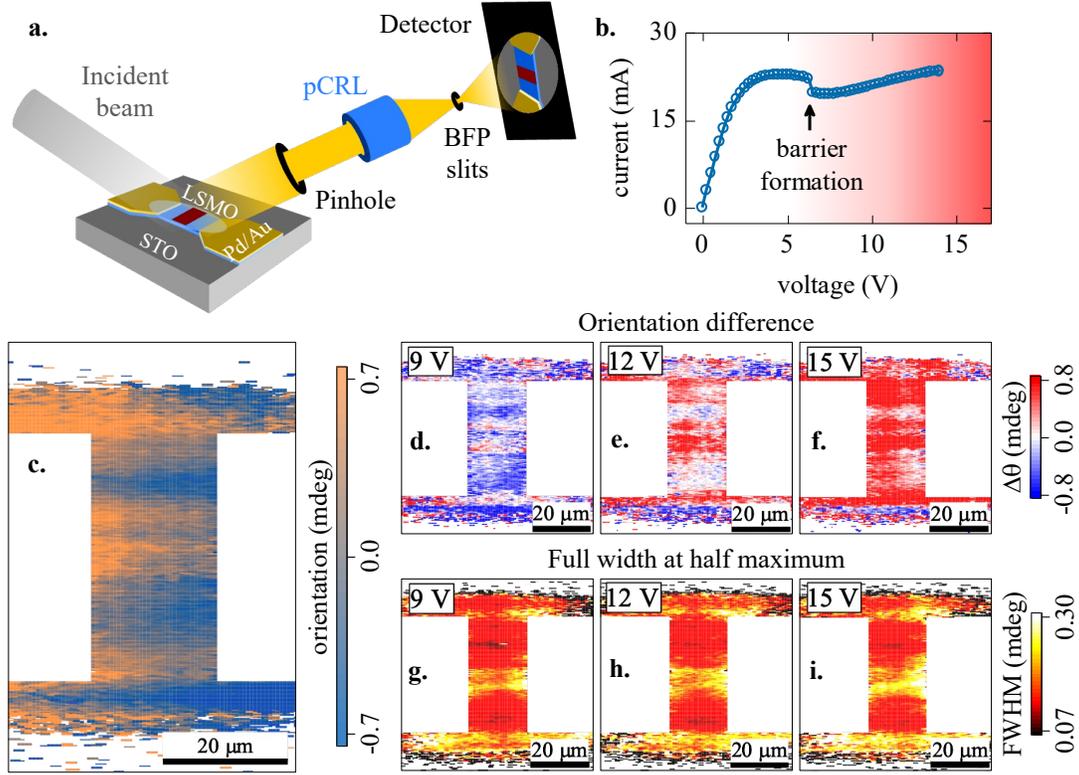

**Fig. 4. a,** Schematic of the DFXM measurement setup. A parallel x-ray beam illuminates the entire device. The diffracted beam passes through a pCRL that refracts the beam producing a magnified device image on the detector. **b,** I-V curve of a 20×40 μm² LSMO device at 115 K. **c,** DFXM rocking curve COM map showing the lattice orientation distribution in the device at 0 V. **d-e,** DFXM rocking curve COM difference maps corresponding to the applied voltages of 9 V (**d**), 12 V (**e**), and 15 V (**f**). These maps show how the lattice orientation changes during the MIT triggering. **g-i,** DFXM rocking curve width maps measured under applied voltages of 9 V (**g**), 12 V (**h**), and 15 V (**i**). These maps show how the rocking curve width changes throughout the device during the MIT triggering. In the **d-i** maps, the formation of a barrier feature in the device center can be observed.

We observed a significant crystal lattice reorientation throughout the entire device when LSMO undergoes the MIT switching. The spatial distribution of this reorientation is consistent with the voltage-induced insulating barrier formation. At 9 V (Fig. 4d), the area in the center of the device, which can be identified as the barrier, tilts toward higher angles (red color), while the crystal lattice in the rest of the device, constituting the metallic matrix, tilts toward opposite low angles (blue color), i.e., both switched and unswitched regions experience lattice tilting. This electrically induced lattice tilting of the metal and insulator regions reaches approximately ±0.8 mdeg. As the voltage is increased to 12 V (Fig. 4e), the area occupied by the positive-angle-tilt region grows, which can be attributed to the barrier expansion. At 15 V (Fig. 4f), the positive-angle-tilt regions become prevalent throughout the device and only small, isolated pockets of the negative angle tilt regions remain.

The maps of the full width at half maximum (FWHM) of the rocking curves (Fig. 4 g-i) provide another perspective on the structural changes in LSMO induced by the MIT triggering. We observed broadening of the rocking curves, indicating the increasing spread of the lattice orientation distribution as the device undergoes the switching. The rocking curve broadening appears to be spatially modulated. Larger broadening, up to ~0.3 mdeg, occurs in the device center, which can be attributed to the voltage-induced insulating barrier formation. The FWHM of the rocking curves in the unswitched, metallic regions near the device edges remains at ~0.15 mdeg. As the applied voltage increases, the center region with the increased FWHM expands, but it does not spread throughout the full device area even at 15 V, unlike in the maps of the lattice orientation distribution discussed in the previous paragraph (Fig. 4 d-f). We can conclude therefore, that similar to the microdiffraction measurements, DFXM shows the development of the lattice distortions across the device as the insulating barrier forms during the MIT switching, but the boundaries of the barrier region are not well defined from the structural perspective.

## IV. Discussion

Previous magneto-optical measurements have shown a sharp contrast between the insulating barrier and metal matrix when an applied voltage triggers the phase transition in LSMO devices [33]. Such clear spatial separation of metal/insulator phase regions is a direct consequence of magneto-optical imaging contrast sensitivity: the material can locally be either in the ferromagnetic metal or paramagnetic insulator phase. The local x-ray diffraction measurements presented in this work (microdiffraction and DFXM) revealed that the boundary between the electrically induced metal/insulator phase separation in LSMO is not structurally well defined. When an insulating barrier focuses Joule heating and causes local lattice thermal expansion, this expansion propagates partially throughout the entire device (Fig. 2a). Mechanical coupling between the hot insulating and cold metallic regions is likely the origin of lattice twinning, i.e., the formation of structural domains of different inclinations, which was evidenced by the (i) reduced diffraction intensity (Fig. 2b), (ii) increased Bragg peak width (Fig. 2c), (iii) Bragg peak splitting (Fig. 3), and (iv) orientation and width changes of the rocking curves (Fig. 4). Because such lattice twinning does not occur during equilibrium thermal expansion, it is not justified to directly compare the voltage-induced lattice expansion and equilibrium thermal expansion in order to estimate the local temperature increase during the MIT switching, i.e., to use strain as a temperature probe. Joule heating in the film also causes thermal expansion in the underlying substrate (gray symbols in Fig. 2e). The substrate expansion, however, is a factor of ~3-5 smaller (depending on the position) compared to the film, which suggests that the $c/a$ ratio in LSMO could be changing when the applied voltage is ramped up. It is well established that the lattice strain and $c/a$ ratio have a major contribution in determining the electrical and magnetic properties in complex oxide thin films [49,50]. Specifically in LSMO, strain can change the phase transition temperature [51], magnetic anisotropy [52–54], magnetotransport [55], magnetization domain configuration [56], *etc*. The voltage-induced inhomogeneous lattice expansion occurring throughout the entire LSMO device, as observed in this work, can play an important role in the process of electrical MIT triggering as well as contribute to the previously reported magnetic anisotropy change and anomalous magnetotransport properties in the LSMO switching devices [34,35].

## V. Conclusions

Local microstructural studies revealed that electrical MIT triggering can be accompanied by the development of substantial inhomogeneous lattice strain, even when the phase transition does not directly involve a coincident 1$^{st}$ order structural transition, such as in the case of LSMO presented in this work. The measured voltage-induced strain exhibited a nonuniform profile across the device length, i.e., larger strain develops in the device center than at the edges. This strain nonuniformity can be attributed to the Joule heating being focused into the regions where LSMO locally undergoes the switching into the insulating phase, resulting in a higher local temperature and, consequently, in a larger thermal expansion. While the Joule-heating-driven lattice expansion is anticipated, local diffraction measurements further revealed unexpected qualitative differences between the electrothermal and equilibrium thermal crystal structure evolution. We found evidence of lattice twinning and development of pronounced lattice distortions during the MIT triggering. The voltage-induced structural changes are not confined only within the local region where the MIT is triggered, but the lattice distortions propagate throughout the device. Microdiffraction experiments showed that the lattice expansion within the entire device area persistently increases as the voltage is being ramped up past the MIT triggering threshold, even though a constant temperature differential between the unswitched metal phase regions and switched insulator phase regions is expected to be established after the insulating barrier forms inside the metallic matrix. DFXM imaging showed that the crystal lattice experiences tilting and orientation distribution spread both in the switched insulating and unswitched metallic regions. The development of the above structural inhomogeneities is likely a consequence of the electrically nonlinear MIT switching process (acting to establish a large thermal contrast between the hot insulating barrier and cold metallic matrix) that is being subjected to mechanical constraints (acting to maintain the film continuity). It is well known that MIT materials are extremely sensitive to even minute crystal structure changes. Our observations of the inhomogeneous strain development during electrical MIT triggering suggest that the switching induced structural changes in the material may play a significant role in the switching process, for example, influencing properties such as the switching threshold voltage/current or the shape and minimum size of the switched region. Furthermore, our observations of local voltage-induced inhomogeneous strain and lattice distortions may provide an unconventional way to explore local non-equilibrium phenomena in MIT systems by electrically driving the material into a structural configuration that differs from the thermal equilibrium state.

## Acknowledgments

This work was supported as part of the "Quantum Materials for Energy Efficient Neuromorphic Computing" (Q-MEEN-C), an Energy Frontier Research Center funded by the U.S. Department of Energy, Office of Science, Basic Energy Sciences under the Award No. DESC0019273. This research used resources of the Advanced Light Source, a U.S. Department of Energy (DOE) Office of Science User Facility under contract no. DE-AC02-05CH11231. This research used resources of the Advanced Photon Source, a U.S. Department of Energy (DOE) Office of Science user facility operated for the DOE Office of Science by Argonne National Laboratory under Contract No. DE-AC02-06CH11357. The authors would like to thank the Karlsruhe Nano Micro Facility (KNMF) for the fabrication of the polymer x-ray optics.

# Local strain inhomogeneities during the electrical triggering of a metal-insulator transition revealed by the x-ray microscopy

*Pavel Salev[1], Elliot Kisiel[2,3], Dayne Sasaki[4], Brandon Gunn[2], Wei He[2], Mingzhen Feng[4], Junjie Li[2], Nobumichi Tamura[5], Ishwor Poudyal[3], Zahir Islam[3], Yayoi Takamura[4], Alex Frano[2], Ivan K. Schuller[2]*

[1]Department of Physics and Astronomy, University of Denver, Denver CO
[2]Department of Physics, University of California San Diego, La Jolla CA
[3]X-ray Science Division, Argonne National Laboratory, Argonne IL
[4]Department of Materials Science and Engineering, University of California Davis, Davis CA
[5]Advanced Light Source, Lawrence Berkeley National Laboratory, Berkeley CA

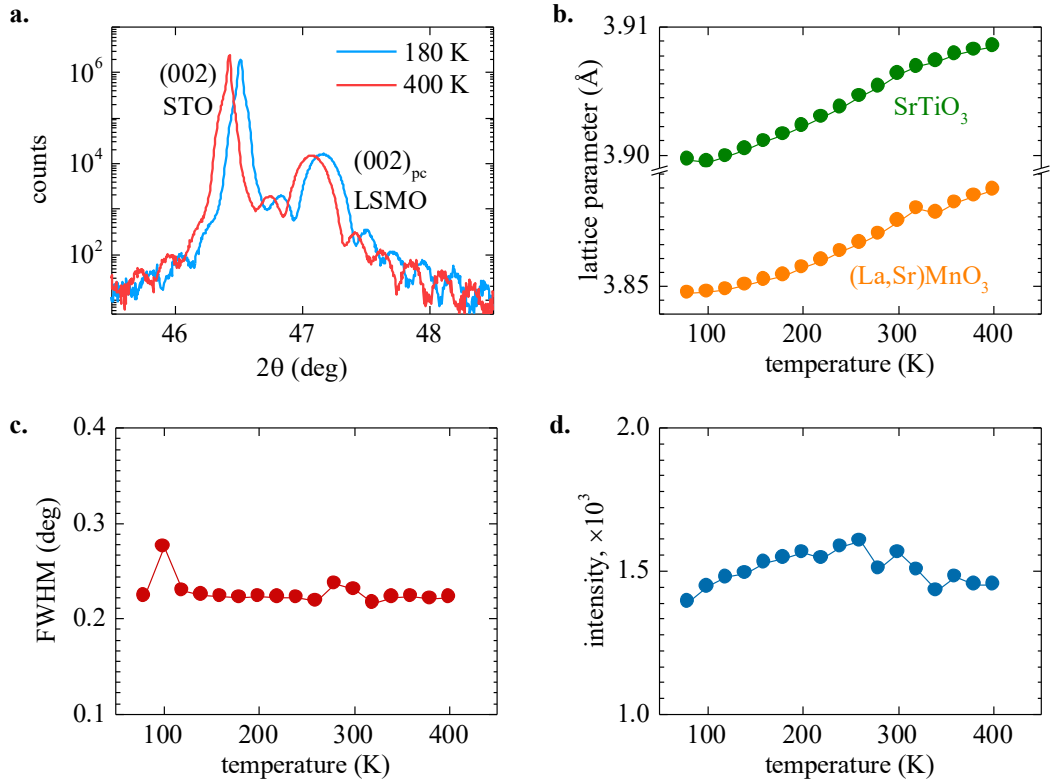

**Suppl. Fig. S1.** Temperature dependent x-ray diffraction measurements of an unpatterned (La,Sr)MnO$_3$ thin film sample. **a,** Diffraction pattern in the vicinity of (002) Bragg peak of the SrTiO$_3$ substrate recorded at two temperatures, 180 K (blue line) and 400 K (red line). Equilibrium thermal expansion of both film and substrate can be observed. **b,** Temperature dependence of the out-of-plane lattice constants of SrTiO$_3$ substrate (green symbols) and (La,Sr)MnO$_3$ film (orange symbols). **c, d,** Temperature dependence of the full-width-at-half-maximum (**c**) and intensity (**d**) of the (002)$_{pc}$ Bragg peak of (La,Sr)MnO$_3$. The curves in both panels show only a weak temperature dependence. All measurements were performed in a laboratory-based x-ray system using monochromatic Cu K-α source.